\documentclass{elsart}

\usepackage{graphicx}
\usepackage{amssymb}

\newcommand{\im}{\mathop{\mathrm{Im}}}
\begin{document}

\begin{frontmatter}

\title{Quantum Hall effect on the Lobachevsky plane}

\author{D.V. Bulaev\corauthref{cor}},
\author{V.A. Geyler},
\author {V.A. Margulis}

\corauth[cor]{Corresponding author. E-mail: bulaevdv@mrsu.ru}
\address{Mordovian State University, 430000 Saransk, Russia}

\begin{abstract}
The Hall conductivity of an electron gas on the surface of constant negative curvature (the
Lobachevsky plane) in the presence of an orthogonal magnetic field is investigated. It is shown
that the effect of the surface curvature is to change the break locations and the plateau widths
in the Hall conductivity. An increase of temperature results in smearing of the steps.
\end{abstract}

\end{frontmatter}

\section{Introduction}
The two-dimensional electron gas (2DEG) in quantized magnetic fields has attracted a lot of
attention in recent years, both experimentally and theoretically. An increasing interest is due
to unique magnetic (the de Haas --- van Alphen effect), transport (the quantum Hall effect and
Shubnikov --- de Haas effect), and optic (the cyclotron resonance) properties. Moreover, 2DEG
systems are of great interest, since they are extensively used in modern electronic devices and
hold much promise as building blocks for future electronic and mechanical nanodevices.

One of the fundamental properties of the 2DEG is the quantum Hall effect (QHE). Although this
effect was discovered and theoretically explained about twenty years ago, the interest to the QHE
is increasing up to now. Several works is devoted to the effect of electron-electron \cite{Lee},
electron-phonon \cite{Blanter}, and spin-orbit \cite{Meir,Crepieux} interactions on the Hall
conductivity. Interesting features are obtained from an analysis of the Hall conductivity at low
magnetic fields \cite{Huckestein} or from disorder effects \cite{Cavalcanti,Giacconi}.

Other branch of the QHE physics both theoretical
\cite{Magarill,Batista,Melik,Haldane,Iengo,Ali96,Ali99,Carey98,Carey99,Grosche,Avron,Pnueli} and
experimental \cite{Ford,Leadbeater} concern the study of the surface curvature effects on the
transport properties. The recent progress in nanotechnology has made it possible to produce
curved 2D layers \cite{Prinz00} and nanometer-size objects of desired shapes \cite{Prinz01}. In
particular, an original technique developed in Refs. \cite{Prinz00,Prinz01} enables fabricating
nanotubes, quantum rolls, rings, and spiral-like strips of precisely controllable shapes and
dimensions. The QHE on different non-flat surfaces have been studied, for example, on the quantum
cylinder \cite{Magarill}, sphere \cite{Batista,Melik}, torus \cite{Haldane}, and surface of
constant negative curvature \cite{Iengo,Ali96,Ali99,Carey98,Carey99,Grosche,Avron,Pnueli}.

In this paper we study the effect of the surface curvature of the 2DEG on the Hall conductivity.
The case of the 2DEG on the surface of constant negative curvature (the Lobachevsky plane) in an
orthogonal magnetic field is considered. Although the Lobachevsky plane is not accessible to the
experimental realization, the problem of the physics on the Lobachevsky plane has a deep relation
with some interesting problems, like the occurrence of the chaos in the surface of negative
curvature \cite{Grosche2,Gutzwiller}, the Berry phase \cite{Albeverio}, and point perturbations
\cite{Bruning} on the Lobachevsky plane. In recent years, the QHE on the Lobachevsky plane is a
subject of current interest. In particular, the Laughlin wavefunctions of the QHE were studied in
Refs.~\cite{Iengo,Ali96,Ali99}. In Ref. \cite{Carey98} the noncommutative geometry models for the
QHE on the Lobachevsky plane were developed. The QHE in the presence of disorder was investigated
in Ref.~\cite{Carey99}. Grosche \cite{Grosche} showed that the QHE at the Aharonov-Bohm flux has
interesting features. In Refs.~\cite{Avron,Pnueli} scattering theory and the Hall conductance of
leaky tori with constant negative curvature were considered.

\section{Density of states}
We consider the case of noninteracting electrons confined to the surface of constant negative
curvature (the Lobachevsky plane) in a magnetic field $\vec{B}$.

The spinless one-particle Hamiltonian of an electron on a two-dimensional Riemann surface $M$ is
given by
\begin{equation}
\label{eq:H_0}
H=\frac{1}{2m^*}g^{-1/2}\left(\frac{\hbar}{\mathrm{i}}\partial_\mu-\frac{e}{c}A_\mu\right)
g^{1/2}g^{\mu\nu}\left(\frac{\hbar}{\mathrm{i}}\partial_\nu-\frac{e}{c}A_\nu\right)+\frac{\hbar^2}{8m^*}
\frac{R}{2},
\end{equation}
where $m^*$ is the effective electron mass, $g^{\mu\nu}$ is the contravariant component of the
metric tensor of the manifold, $g=\det g_{\mu\nu}$, $A_\mu$ is the component of the vector
potential of a magnetic field $\vec{B}$, the last term in Eq.~(\ref{eq:H_0})  is the surface
potential which arises from the surface curvature \cite{Lan}.

We shall employ the Poincar\'e realization in which the Lobachevsky plane $M$ is identified with
the upper complex halfplane $M=\{z=x+\mathrm{i}y\in \Cset:\ y>0\}$ endowed with the metric
\[
ds^2=\frac{a^2}{y^2}(\d x^2+\d y^2),
\]
where $a$ is the radius of curvature. Therefore, in the Landau gauge ($\vec{A}=(Ba^2y^{-1},0)$),
the Hamiltonian (\ref{eq:H_0}) on the Lobachevsky plane is given by
\begin{equation}
\label{eq:Hamiltonian}
H=\frac{\hbar^2}{2m^*a^2}\left[-y^2\left(\partial_x^2+
\partial_y^2\right)+2\mathrm{i}by\partial_x+b^2-\frac14\right],
\end{equation}
where $b=eBa^2/\hbar c$. The spectrum of $H$ consists of two parts \cite{Comtet}: a point
spectrum in the interval $(0,\hbar^2b^2/2m^*a^2)$ consisting of a finite number of Landau levels
\begin{equation}
\label{eq:Energy}
E_n=\hbar\omega_c\left(n+\frac12\right)-\frac{\hbar^2}{2m^*a^2} \left(n+\frac12\right)^2,\ 0\le
n<|b|-\frac12
\end{equation}
and an absolutely continuous spectrum in the interval $[\hbar^2b^2/2m^*a^2,\infty)$
\[
E(\nu)=\frac{\hbar^2}{2m^*a^2}\left(b^2+\nu^2\right),\ 0\le\nu<\infty.
\]

The electron density of states (DOS) $n(E)$ per unit area is defined by the following expression:
\[
n(E)=\frac{1}{S}\int \im G(\vec{r},\vec{r};E+\mathrm{i}0)\d\vec{r},
\]
where $S$ is the area of the surface and $G(\vec{r},\vec{r'};E)$ is the Green's function of the
Hamiltonian. In the case of homogeneous systems, the renormalized Green's function
$G^{\mathrm{ren}}(\vec{r},\vec{r};E)$ coincide with the so-called Krein's function $Q(E)$, which
for the Lobachevsky plane is given by \cite{Bruning}
\[
Q(E)=-\frac{m^*}{2\pi\hbar^2}\left[\psi(t-b)+\psi(t+b)+2\gamma_\mathrm{E}-\ln 4a^2 \right],
\]
where $\psi(z)=[\ln\Gamma(z)]'$, $t(E)=1/2+\sqrt{b^2-2m^*a^2 E/\hbar^2}$, and $\gamma_\mathrm{E}$
is the Euler number. By applying the properties of $\psi$-function and the Sochocki formula
$\delta(x)=-(1/\pi)\im(1/(x+\mathrm{i}0))$, one obtains the following expression for the electron
density of states:
\begin{eqnarray}\nonumber
n(E)&=&\frac{1}{2\pi a^2}\sum_{0\le n<|b|-1/2}
\left(|b|-n-\frac12\right)\delta(E-E_n)\\
\label{eq:DOS}&&+\frac{m^*}{2\pi\hbar^2}
\Theta\left(E-\frac{\hbar^2b^2}{2m^*a^2}\right) \frac{\sinh 2\pi \sqrt{2m^*a^2E/\hbar^2-b^2}}
{\cosh 2\pi\sqrt{2m^*a^2E/\hbar^2-b^2}+\cos 2\pi b},
\end{eqnarray}
where $\Theta(x)$ is the Heaviside step function. The first term in Eq.~(\ref{eq:DOS})
corresponds to the point spectrum and the second term corresponds to the continuous one and
coincides with the expression given in Ref.~\cite{Comtet}.

\begin{figure}[!htp]
\begin{center}
\includegraphics[width=12cm]{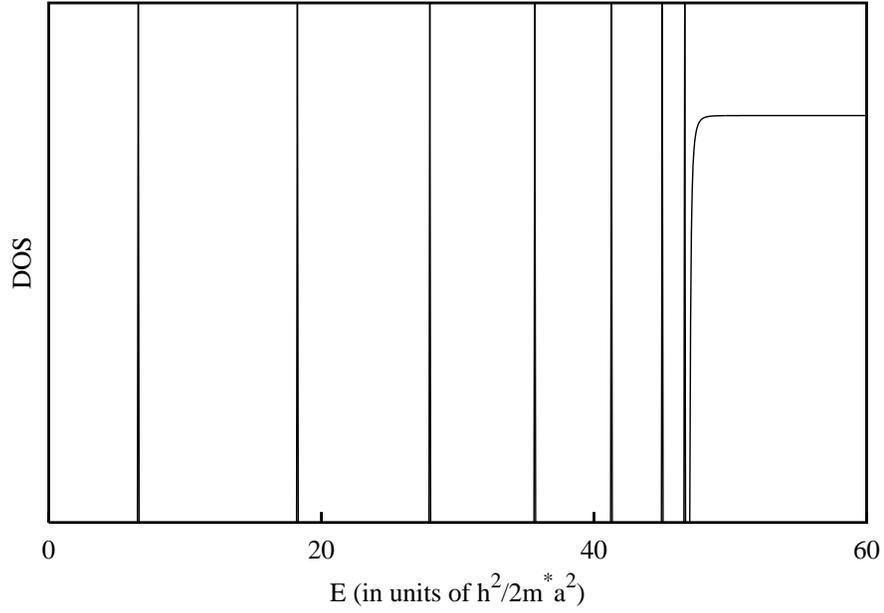}
\end{center}
 \caption{Density
of states on the Lobachevsky plane as a function of the energy; $b=7$, $a=3\times10^{-6}\;$cm.}
\label{fig:1}
\end{figure}

\begin{figure}[!htp]
\begin{center}
\includegraphics[width=12cm]{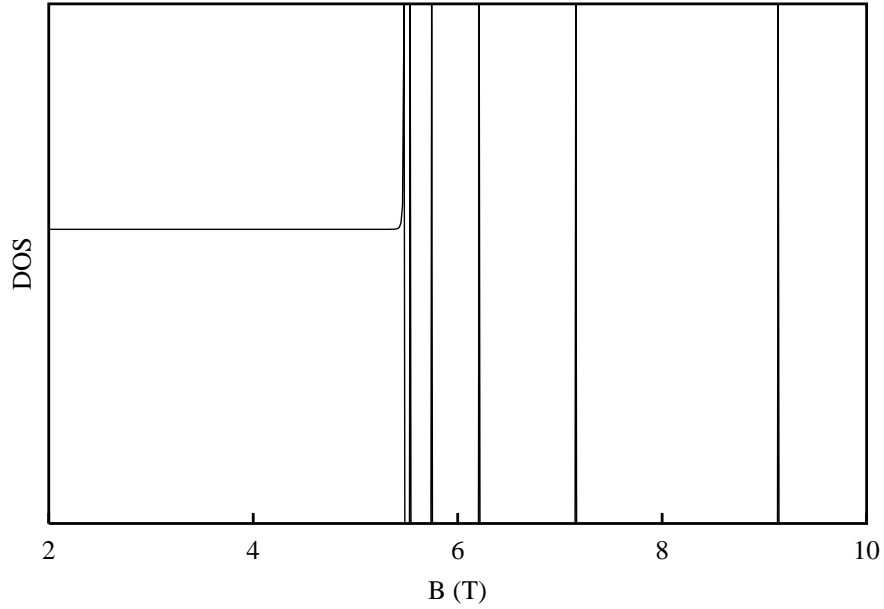}
\end{center}
 \caption{Density
of states on the Lobachevsky plane as a function of a magnetic field; $2m^*a^2E/\hbar^2=65.25\;$,
$a=3\times10^{-6}\;$cm. At this energy the threshold of the step take place at $b=7.5$.}
\label{fig:2}
\end{figure}

In Figures~\ref{fig:1} and \ref{fig:2} we plot the dependencies of the DOS on the energy and on
the magnetic field, respectively. In these figures we show schematically the delta-peaks,
corresponding to the discrete spectrum. The step-like dependence of the density of states
correspond to the continuous spectrum. As can be seen from Eq.~(\ref{eq:DOS}), for
$E=\hbar^2b^2/2m^*a^2$, on the plateau of the step, the DOS approaches $m^*/2\pi\hbar^2$
asymptotically with increasing energy or with decreasing magnetic field. Note that the step is a
smeared one if $b$ is close to an integer (see Fig.~\ref{fig:1}). If $b$ is close to a
half-integer, then the sharp peak appear at the threshold of the step (see Fig.~\ref{fig:2}). The
appearance of this peak is defined as follows. For half-integer $b$ and $b^2=2m^*a^2E/\hbar^2$
(i.e. at the threshold of the step), the denominator of the second term in Eq.~(\ref{eq:DOS}) is
zero and the infinite peak appears (if $b$ is close to the a half-integer, then the peak height
is finite).

Note that in the limit of zero curvature  ($a\to\infty$), we get the well known formula for the
DOS on the flat surface:
\[
n(E)=\frac{|Be|}{2\pi\hbar c}\sum_{n=0}^\infty\delta(E-E_n).
\]

In this work we consider the case of high magnetic fields and the large radius of curvature
($a^2B^2e^2/2m^*c^2>E_{\mathrm{F}}$). In this case, the energy spectrum below the Fermi energy $E_{\mathrm{F}}$ is
discrete one only. Therefore, the second term in Eq.~(\ref{eq:DOS}) is zero.

\section{Hall conductivity}

In the linear response approximation, Str\v{e}da  \cite{Streda} has shown that the Hall
conductivity is given by the following expression, when the Fermi energy is in an energy gap:
\begin{equation}
\label{eq:sigma}
\sigma_{xy}(E_{\mathrm{F}},0)=\frac{ec}{S}\frac{\partial N}{\partial B},
\end{equation}
where $N$ is the number of states below the Fermi energy. In the case of large radius of
curvature and high magnetic fields $(a^2B^2e^2/2m^*c^2>E_{\mathrm{F}})$, using Eq.~(\ref{eq:DOS}) we obtain
\begin{eqnarray}
\nonumber N&=&S\int_{-\infty}^{E_{\mathrm{F}}}n(E)\d E=\frac{S}{2\pi a^2}\left[b+\frac12-\sqrt{b^2-2m^*a^2 E_{\mathrm{F}}/\hbar^2}\right]\\
\label{eq:N}
&&\times\left(b-\frac12 \left[b+\frac12-\sqrt{b^2-2m^*a^2 E_{\mathrm{F}}/\hbar^2}\right]\right),
\end{eqnarray}
where $[x]$ is the integer part of $x$ (we consider for simplicity the case of $b>0$ only).

It is easy to see that in an energy gap the integer part of $b+1/2-\sqrt{b^2-2m^*a^2
E_{\mathrm{F}}/\hbar^2}$ is constant. Therefore, substituting Eq.~(\ref{eq:N}) into
Eq.~(\ref{eq:sigma}), we obtain
\begin{equation}
\label{eq:sigma_0}
\frac{\sigma_{xy}(E_{\mathrm{F}},0)}{\sigma_0}= -\left[b+\frac12-\sqrt{b^2-2m^*a^2
E_{\mathrm{F}}/\hbar^2}\right],
\end{equation}
where $\sigma_0=e^2/h$.

As can be seen from Eq.~(\ref{eq:sigma_0}), the field dependence of the Hall conductivity has a
step-like structure. In the limit of zero curvature  ($a\to\infty$), we get the well known
formula for the Hall conductivity on the flat surface:
\[
\frac{\sigma_{xy}(E_{\mathrm{F}},0)}{\sigma_0}\mathop{\longrightarrow}\limits_{a\to\infty}
-\left[\frac12+\frac{E_{\mathrm{F}}}{\hbar\omega_c}\right].
\]

\begin{figure}[!htp]
\begin{center}
\includegraphics[width=12cm]{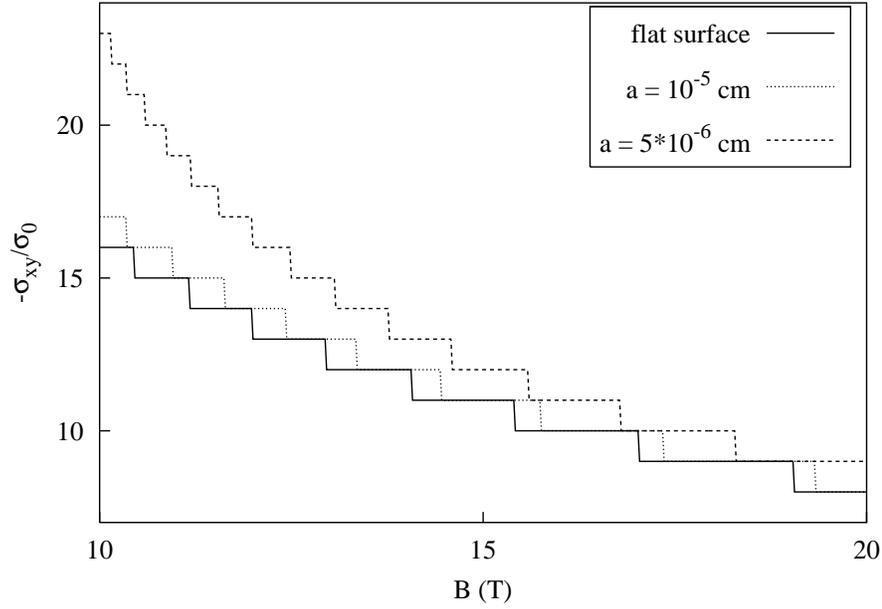}
\end{center}
 \caption{Hall
conductivity as a function of a magnetic field; $T= 0\;$K, $E_{\mathrm{F}}= 5\times10^{-13}\;$erg.}
\label{fig:3}
\end{figure}

The breaks in the conductivity arise from the crossings of the Fermi energy by Landau levels.
Therefore, the break locations are defined by
\begin{equation}
\label{eq:jump}
E_{\mathrm{F}}=\hbar\omega_c\left(n_0-\frac12\right)-\frac{\hbar^2}{2m^*a^2} \left(n_0-\frac12\right)^2,
\end{equation}
where $n_0$ is the number of fully occupied Landau levels below the Fermi energy. As can be seen
from this equation, the effect of the surface curvature  is to shift the break locations to
higher magnetic fields. The shift of the break location is equal to $\Phi_0(n_0-1/2)/4\pi a^2$
(see Fig.~\ref{fig:3}).

Note that $\sigma_{xy}(E_{\mathrm{F}},0)=-\sigma_0 n_0$ on the plateaus. From Eq.~(\ref{eq:jump}) we find
the plateau width
\begin{equation}
\label{eq:dB} \Delta
B=\frac{m^*c}{|e|\hbar}\left(\frac{E_{\mathrm{F}}}{n_0^2-1/4}-\frac{\hbar^2}{ 2m^*a^2}\right).
\end{equation}
Thus the plateau width for the Lobachevsky plane less than for the flat surface by $\Phi_0/4\pi
a^2$, where $\Phi_0=hc/|e|$ is the magnetic flux quantum.

\begin{figure}[!htp]
\begin{center}
\includegraphics[width=12cm]{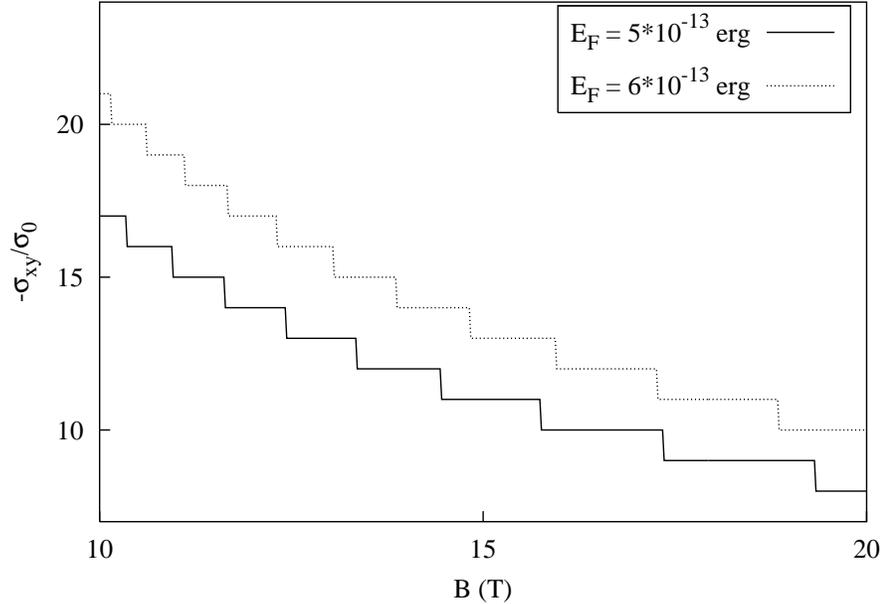}
\end{center}
 \caption{Hall
conductivity as a function of a magnetic field; $T= 0\;$K, $a= 10^{-5}\;$cm.} \label{fig:4}
\end{figure}

In Fig.~\ref{fig:4} we plot $\sigma_{xy}(B)$ at different $E_{\mathrm{F}}$. It can be seen that the plateau
width increases with increasing the Fermi energy and the break locations are shifted to higher
magnetic fields.

Let us consider the influence of temperature on the Hall conductivity. The dependence of
$\sigma_{xy}$ on temperature is given by
\[
\sigma_{xy}(\mu,T)=\int\limits_{-\infty}^\infty\left(-\frac{\partial f_0(E)}{\partial E}
\right)\sigma_{xy}(E,0)\d E,
\]
where $\mu$ is the chemical potential, $f_0(E)$ is the Fermi function, and $\sigma_{xy}(E,0)$ is
the Hall conductivity at zero temperature. In the case of strong magnetic quantization, we can
neglect the contribution of electrons with the energies $E\gg E_{\mathrm{F}}$ lying in the continuous
spectrum  to the Hall conductivity. Therefore,
\begin{equation}
\label{eq:sigma_T}
\frac{\sigma_{xy}(\mu,T)}{\sigma_0}=-\sum_{n=0}^{[b-1/2]}f_0(E_n)+\frac{\left[b-1/2\right]}{
1+\exp\{(\hbar^2b^2/2m^*a^2-\mu)/T\}}.
\end{equation}

\begin{figure}[!htp]
\begin{center}
\includegraphics[width=12cm]{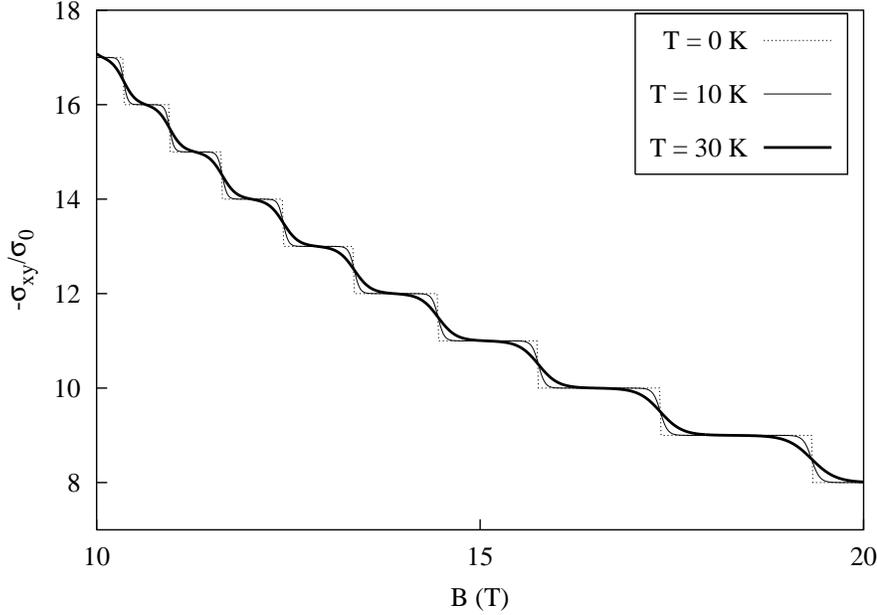}
\end{center}
 \caption{Hall
conductivity as a function of a magnetic field; $\mu= 5\times10^{-13}\;$erg, $a= 10^{-5}\;$cm.}
\label{fig:5}
\end{figure}

As shown in Fig.~\ref{fig:5}, an increase of temperature results in smearing of the steps. The
smearing is essential for the steps with smaller plateau width.

\begin{figure}[!htp]
\begin{center}
\includegraphics[width=12cm]{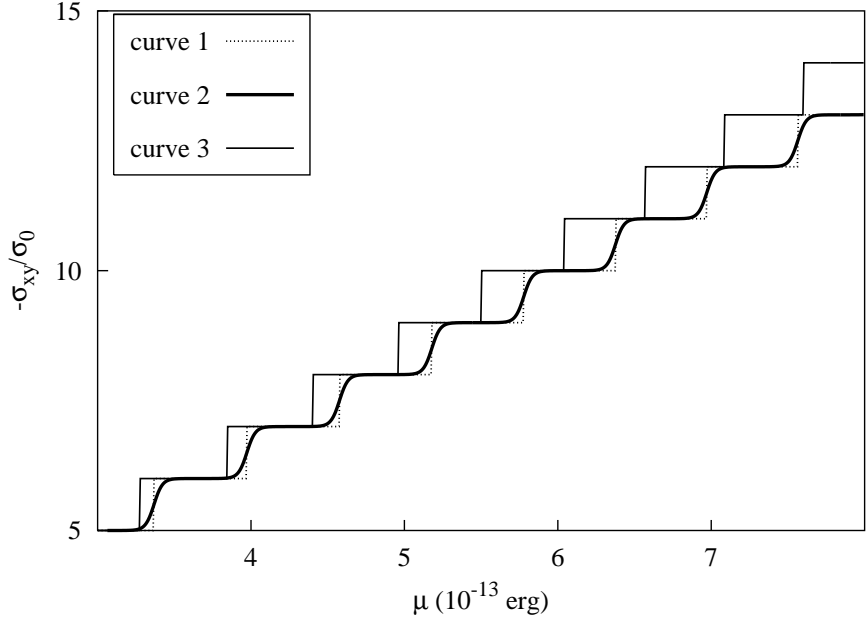}
\end{center}
 \caption{Hall
conductivity as a function of the chemical potential in the magnetic field $B=20\;$T; curve~1:
$T=0\;$K, $a=10^{-5}\;$cm; curve~2: $T=20\;$K, $a=10^{-5}\;$cm; curve~3: $T=0\;$K,
$a=5\times10^{-6}\;$cm.} \label{fig:6}
\end{figure}

Let us consider the dependence of the Hall conductivity on the chemical potential. From
Eq.~(\ref{eq:jump}) we find that the effect of the surface curvature  is to shift the break
locations to lower values of chemical potential. This shift of the break location is equal to
$\hbar^2(n_0-1/2)^2/2m^*a^2$ (see Fig.~\ref{fig:6}).

The plateau width is given by
\begin{equation}
\label{eq:dmu} \Delta\mu\vert_{T=0}=\hbar\omega_c-
\frac{\hbar^2}{m^*a^2}n_0.
\end{equation}
Therefore, the plateau width for the Lobachevsky plane less than for the flat surface by the
value proportional to the number of fully occupied Landau levels.

\section{Conclusions}
In conclusion, we have studied the effect of the surface curvature on the Hall conductivity. The
case of constant negative curvature (the Lobachevsky plane) in an orthogonal magnetic field have
been investigated. It has been shown that the effect of the surface curvature is to change the
break locations and the plateau widths; namely, the surface curvature shifts the break locations
to higher values of magnetic fields (to lower values of the chemical potential) in the dependence
$\sigma_{xy}(B)$ ($\sigma_{xy}(\mu)$). Note that  the shift of break locations are increasing
with increasing the number of fully occupied Landau levels below the Fermi energy. In the
dependence of $\sigma_{xy}$ on $B$, the plateau width for the Lobachevsky plane less than for the
flat surface by $\Phi_0/4\pi a^2$ (see Eq.~(\ref{eq:dB})). In the dependence of $\sigma_{xy}$ on
$\mu$, the plateau width is defined by Eq.~(\ref{eq:dmu}). As can be seen from this equation,
curvature decreases the plateau width. Moreover, the plateau width for the Lobachevsky plane less
than for the flat surface by the value proportional to the number of fully occupied Landau levels
(see Fig.~\ref{fig:6}). An increase of temperature results in smearing of the steps. The smearing
is essential for the steps with smaller plateau width (see Fig.~\ref{fig:5}).

\section*{Acknowledgements}
This work was supported by the INTAS (Grant No.~00-257) and the Russian Ministry of Education
(Grant No.~E02-3.4-370). One of us (V.A.G.) was supported in part by the RFBR (Grant
No.~02-01-00804-a) and the DFG (Grant No.~436 RUS 113/572/0-2).

\end{document}